\documentclass[preprint,12pt]{elsarticle}

\usepackage{amssymb}
\usepackage{subfigure}
\usepackage{hyperref}
\hypersetup{hypertex=true,
	colorlinks=true,
	linkcolor=blue,
	anchorcolor=blue,
	citecolor=blue}
\usepackage{amsmath}
\usepackage{float}
\usepackage{lineno}
\usepackage{caption}
\usepackage{graphicx}
\usepackage{booktabs}
\DeclareCaptionLabelFormat{bold}{\textbf{Fig. #2}}
\DeclareCaptionLabelFormat{bold2}{\textbf{Tab. #2}}
\usepackage{titlesec}
\titleformat{\subsection}{\bfseries}{\thesubsection}{1em}{}
\usepackage{xcolor}

\journal{Journal of Radioanalytical and Nuclear Chemistry}

\begin{document}

\begin{frontmatter}

\biboptions{sort&compress}

\title{Study on the radon adsorption capability of low-background activated carbon}

\author[a]{Chi Li}

\author[b,c,d]{Yongpeng Zhang}

\author[a]{Lidan Lv}

\author[b,c,d]{Jinchang Liu}

\author[b,c,d]{Cong Guo\corref{mycorrespondingauthor}}
\cortext[mycorrespondingauthor]{Corresponding authors}
\ead{guocong@ihep.ac.cn}

\author[b,c,d]{Changgen Yang}

\author[a]{Tingyu Guan}

\author[a]{Yu Liu}

\author[e]{Yu Lei}

\author[a]{Quan Tang\corref{mycorrespondingauthor}}
\ead{tangquan528@sina.com}

\address[a]{School of Nuclear Science and Technology, University of South China, Hengyang, 421001, China
}
\address[b]{Experimental Physics Division, Institute of High Energy Physics, Chinese Academy of Sciences, Beijing,100049, China}
\address[c]{School of Physics, University of Chinese Academy of Sciences, Beijing, 100049, China}
\address[d]{State Key Laboratory of Particle Detection and Electronics, Beijing, 100049, China}
\address[e]{School of Electronic, Electrical Engineering and Physics, Fujian University of Technology, Fuzhou, 350118, China}

\begin{highlights}

\item Developing a radon emanation measurement system with a sensitivity of 73~$\mu$Bq/sample;
\item Identifying a new kind of low-background activated carbon with high radon adsorption efficiency using a self-developed system;
\item Establishing the relationship between radon adsorption capacity and average pore diameter of activated carbon.

\end{highlights}

\begin{abstract}
{Radon is a significant background source in rare event detection experiments. Activated Carbon (AC) adsorption is widely used for effective radon removal. The selection of AC considers its adsorption capacity and radioactive background. In this study, using self-developed devices, we screened and identified a new kind of low-background AC from Qingdao Inaf Technology Company that has very high Radon adsorption capacity. By adjusting the average pore size to 2.3 nm, this AC demonstrates a radon adsorption capacity of 2.6 or 4.7 times higher than Saratech or Carboact activated carbon under the same conditions.}

\end{abstract}



\begin{keyword}
Low-background experiment\sep Radon \sep Activated carbon \sep Adsorption capability\sep Radio-active background


\end{keyword}

\end{frontmatter}





\section*{Introduction}
Rare event detection~\cite{r1,r2,r3,r4} is one of the most cutting-edge directions in modern fundamental physics that holds great potential for significant scientific discoveries.  However, detecting rare event in experiments poses significant challenges due to their extremely low cross-sections with baryonic matter. Although there are various types of rare event searching detectors such as semiconductor detectors~\citep{r5}, nobel element detectors~\citep{r6,r7}, crystal detectors~\citep{r8}, deep cryogenic calorimeters~\citep{r9}, they all have extremely stringent requirements on detector background. One of the most crucial background sources is $^{222}$Rn and its short-lived decay products, which can emit $\alpha$, $\beta$, or $\gamma$ particles over a wide energy range. $^{222}$Rn is continuously released from detector materials~\citep{r18} and can diffuse throughout the entire detector system. Therefore, effectively managing and reducing the presence of $^{222}$Rn is essential to minimize background interference in rare event detection experiments.

Activated carbon is a widely used $^{222}$Rn adsorbent and is often used for radon enrichment or radon removal~\citep{r10,r11,r12,r15}. In terms of highly sensitive radon detectors, activated carbon radon enrichment devices can substantially improve the sensitivity of the detectors. At present, the most sensitive device for direct measurement of radon gas is the electrostatic collection radon detector developed by Y. Nakano with a one-day measurement sensitivity of 0.54~mBq/m$^3$~\citep{r22}, whereas several experiments have realized 10~$\mu$Bq/m$^3$ level radon measurements by using low-temperature activated carbon radon enrichment devices~\citep{r31,r10,r32}. In low-background experiments, the radon adsorption capability and the intrinsic background of the activated carbon are two primary considerations. Saratech activated carbon, known for its excellent radon adsorption capability, low background, and affordable price~\citep{r19}, was a preferred choice for radon removal and enrichment in low-background experiments. However, it has been discontinued since $\sim$2018~\citep{r19}. Carboact activated carbon, another widely used option in low-background experiments, boasts the lowest background among all known activated carbons but comes at a significantly higher cost~\citep{r21}. Therefore, an activated carbon with strong radon adsorption capacity, a comparable background to Saratech or Carboact, and a similar price to Saratech is essential for radon-related background study in low-background experiments.

In this paper, we reported a new kind of low-background activated carbon and studied its radon adsorption capability under various conditions. This paper is organized as follows. Sec.2 describes the background screening of activated carbon. Sec.3 introduces the radon adsorption capacity measurement of activated carbon at different conditions. Sec.4 is the conclusion.

\section*{Activated carbon background measurement}
\subsection*{Measurement system}
\label{emanantionsystem}

\begin{figure}[htb]
	\centering
	\includegraphics[width=0.9\linewidth]{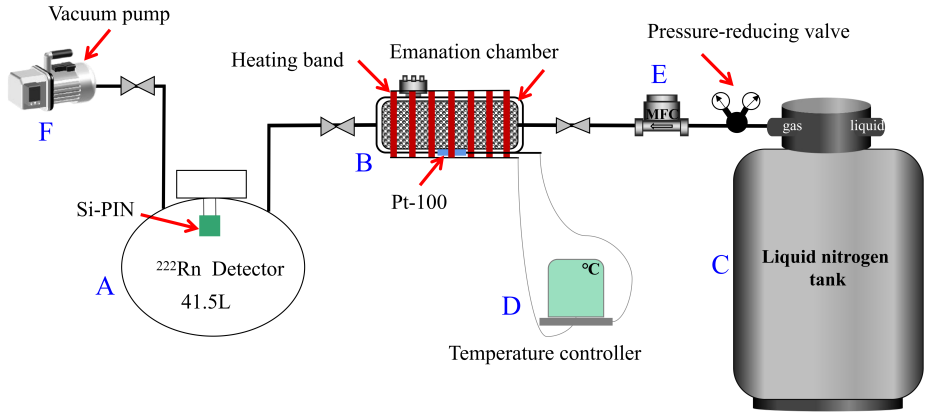}
    \captionsetup{labelsep=period}
	\caption{Radon emanation measurement system}
	\label{Rnemanation}
\end{figure}

The intrinsic background ($^{226}$Ra) of the activated carbon is determined using the radon emanation measurement system shown in Fig.~\ref{Rnemanation}. In contrast to the device reported in Ref~\citep{r33}, we separate the detector chamber from the precipitation chamber, which avoids the influence of the measured material on the detection efficiency of the detector. Different from the device in Ref~\citep{r34}, we used evaporated nitrogen as a carrier gas to blow the radon gas into the detector, which can realize the high radon transport efficiency, in addition, we added a heating device to realize radon desorption from activated carbon. The system used in this work consists of a highly sensitive radon detector, an emanation chamber, a liquid nitrogen tank, a temperature control system, a vacuum pump, a Mass Flow Controler (MFC), and associated pipes and valves.

\begin{figure}[h]
	\centering
	\subfigure{\includegraphics[width=0.45\linewidth]{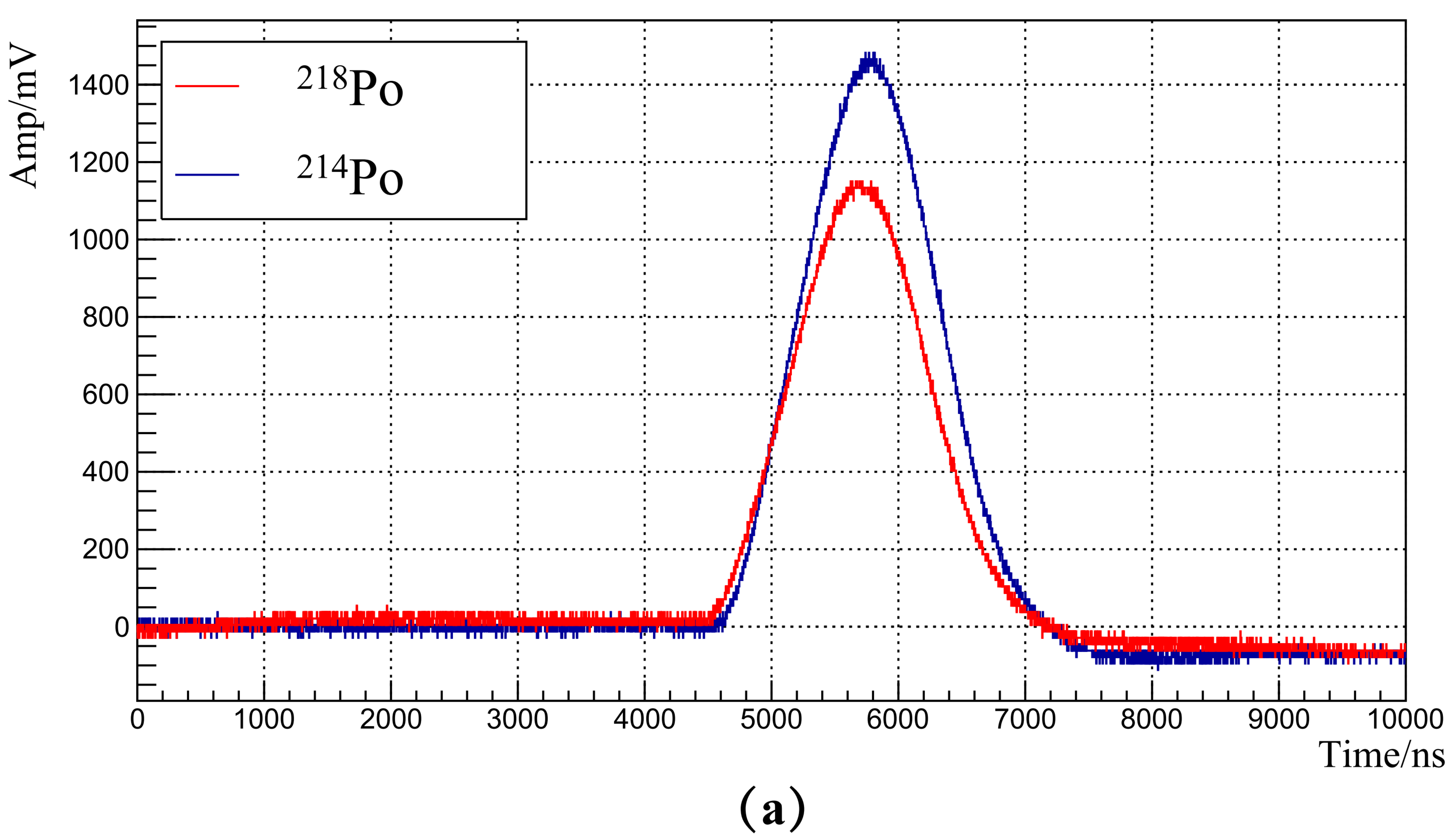}}
	\label{fig:boxingtu}
	\subfigure{\includegraphics[width=0.46\linewidth]{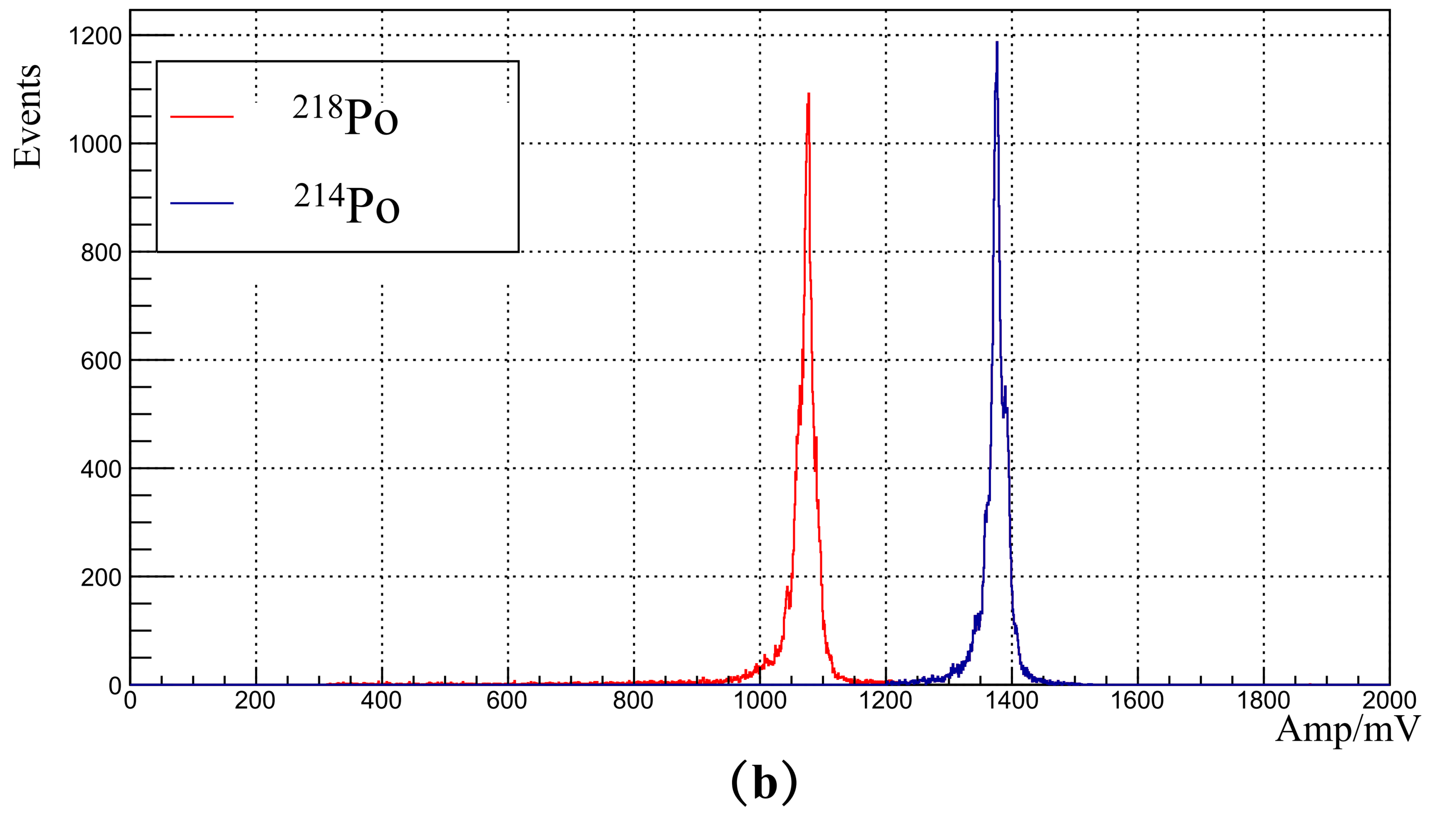}}
	\label{fig:fudutu2}
	\caption{An example pulses of $^{218}$Po and $^{214}$Po signal  and spectrum. (a) The pulses of $^{218}$Po and $^{214}$Po signal. (b) A typical amplitude spectrum of $^{218}$Po and $^{214}$Po signal. }
	\label{tu1}
\end{figure}

(A) The highly sensitive radon detector is employed to measure the radon concentration in the gas. The background event rate of the detector itself is 0.70 $\pm$ 0.15 Counts Per Day (CPD), corresponding to a sensitivity of 0.71~mBq/m$^3$ for a one-day measurement. Example signal pulses are illustrated on the left side of Fig.~\ref{tu1}, while the measured spectrum is displayed on the right side of Fig.~\ref{tu1}. For further information regarding the detector, please refer to Ref~\citep{r13}.

(B) The emanation chamber serves as an activated carbon container. It has a volume of $\sim$4.5~L. A CF35 flange located on the side of the chamber is utilized for inserting the activated carbon. Additionally, two 1/4 VCR valves are installed at both ends of the chamber to regulate the airflow. To prevent the activated carbon from being blown out by the gas flow, stainless steel sintering meshes with an aperture size of 60~$\mu$m are welded at both ends of the chamber.

(C) The liquid nitrogen tank is used to provide evaporating nitrogen for the system. The piping is connected to the vent of the tank. The radon concentration in the evaporating nitrogen is $\sim$0.1 mBq/m$^3$, which is negligible in this measurement.

(D) The temperature control system comprises a temperature controller, a temperature sensor (Pt100), and a heating belt. The heating belt is wrapped around the emanation chamber to provide controlled heating, while the temperature sensor is positioned between the belt and the chamber to serve as the input of the temperature controller.

(E) The gas flow rate in the system is controlled by the Mass Flow Controller (MFC, 1179A, MKS). It ensures precise and accurate control of the gas flow rate within the system.

(F) The vacuum pump (ACP40, Pfeiffer) is used to vacuum the detector before each measurement.

To minimize the background caused by air leakage, the system utilizes knife-edge flanges with metal gaskets and VCR connectors with metal gaskets. The leakage rate of the system is better than 1 $\times$ 10 $^{-8}$~Pa$\cdot$m$^3$/s, which is determined by a helium leak detector (ZQJ-3000, KYKY Technology Co. Ltd).

\subsection*{System sensitivity and measurement results}

The background event rate for the entire emanation system, encompassing the radon detector, emanation chamber, pipelines, and evaporated nitrogen, is 3.00 $\pm$ 1.73 CPD, which corresponds to a sensitivity of 73.2 $\pm$ 7.3~$\mu$Bq/sample. The sensitivity is calculated according to Eq.~\ref{sensitivity}~\citep{r22}.

\begin{equation} 
	L =\frac{1.64 \times \sigma_{BG} \times V_D}{24 \times C_F} 
\label{sensitivity}
\end{equation}

Where L is the sensitivity (90\% confidence level) in the unit of Bq/sample, $\sigma_{BG}$ is the statistical uncertainty of the background event rate which is 1.73 CPD, V$_D$ is the volume of the radon detector which is 0.0415~m$^3$, C$_F$ is the calibration factor of the radon detector which is 67.0 $\pm$ 6.7 Counts Per Hour/(Bq/m$^3$) (CPH/(Bq/m$^3$)). The uncertainty of C$_F$ comes mainly from the instability of the radon source used in the detector calibration~\citep{r13}, and the uncertainty of L is mainly transferred from the C$_F$.

The measurement procedures are as below.

(A) Place the activated carbon sample into the emanation chamber and heat the activated carbon to a temperature of 180$^{\circ}$C.

(B) Purge the system with evaporated nitrogen for 2 hours at a flow rate of 2~L/min. Throughout the purging process, the activated carbon is maintained at a temperature of 180$^{\circ}$C to ensure complete desorption of radon previously adsorbed.

(C) Seal the detector and the emanation chamber for several days to allow $^{226}$Ra to decay into $^{222}$Rn.

(D) Vacuum the detector with the vacuum pump and close the relevant valves after the pressure inside the detector reaches -101~kPa.

(E) Heat the activated carbon to 180$^{\circ}$C and maintain this temperature for at least 30 minutes. This step ensures that any radon gas adsorbed by the activated carbon is fully desorbed~\citep{r10}.

(F) Open the appropriate valves and use evaporated nitrogen to purge the radon gas desorbed from the activated carbon into the radon detector.

(G) Seal the detector for measurement after the pressure inside the detector reaches 0~kPa.

After getting the $^{214}$Po event rate, the $^{226}$Ra concentration in the activated carbon is calculated according to Eq.~\ref{CRa}.

\begin{equation} 
	C_{Ra} =\frac{(n-n_b) \times (V_D+V_E)}{m \times C_F \times 24 \times (1 - e^{\lambda_{Rn} \times t_s}) \times 1000} 
\label{CRa}
\end{equation}

Where C$_{Ra}$ is the $^{226}$Ra concentration in activated carbon in the unit of mBq/kg, n is the event rate for sample measurement in the unit of CPD, n$_b$ is the background event rate in the unit of CPD, V$_D$ is the detector volume in the unit of m$^3$, V$_E$ is the volume of the emanation chamber in the unit of m$^3$, m is the mass of the sample in the unit of kg, C$_F$ is the calibration factor of the detector which is the same as in Eq.~\ref{sensitivity}, $\lambda_{Rn}$ is the $^{222}$Rn decay constant, t$_s$ is the sample sealing time. The uncertainty of C$_R$ consists of both systematic and statistical uncertainties, the systematic uncertainty is transferred from C$_F$, and the statistical uncertainty is transferred from n and n$_b$.

\begin{figure}[htb]
	\centering
	\includegraphics[width=0.8\linewidth]{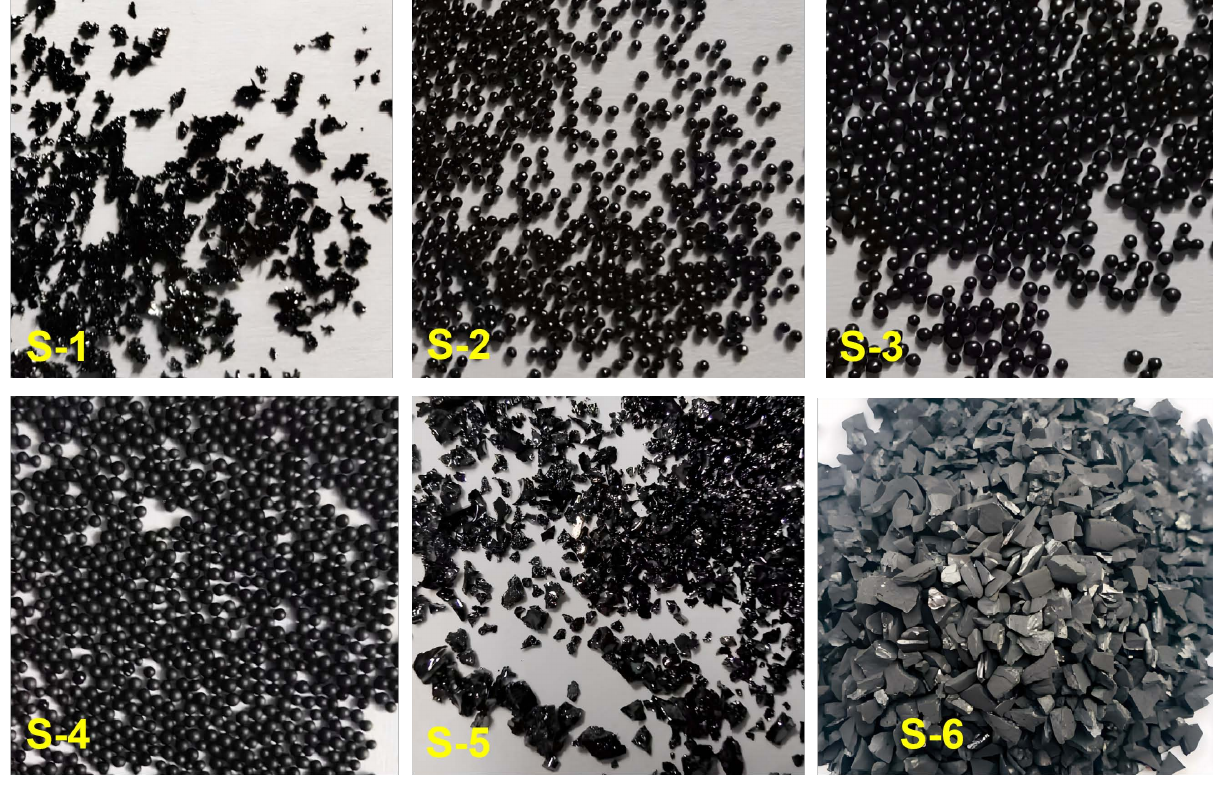}
	\caption{Pictures of the six kinds of activated carbons investigated. S-1 is Carboact activated carbon, S-2 is Saratech activated carbon, S-3 and S-4 are bought from Qingdao Inaf Technology Company (QITC), S-5 is developed by China University of Petroleum (CUP), and S-6 is bought from Shanghai Activated Carbon Company (SACC).} 
	\label{ACpicture}
\end{figure}

Fig.~\ref{ACpicture} presents images of the six types of activated carbon investigated. S-1 is Carboact activated carbon, S-2 is Saratech activated carbon, S-3 and S-4 were obtained from Qingdao Inaf Technology Company (QITC),  S-5 was developed by China University of Petroleum (CUP), and S-6 was purchased from Shanghai Activated Carbon Company (SACC). Except for S-4, the remaining five types of activated carbon are commercially available products. The raw materials of S-1 and S-2 are unknown to us. S-3 and S-4 are made from asphalt but with different technologies, S-5 is made from phenolic resin, and S-6 is made from coconut shells. Although S-3 and S-4 exhibit a similar appearance to S-2, S-2 is significantly harder in texture.

Using the system shown in Fig.~\ref{Rnemanation}, the $^{226}$Ra concentration was measured for the four types of activated carbon illustrated in Fig.~\ref{ACpicture}. Tab.~\ref{background} provides a summary of the $^{226}$Ra concentrations observed in different activated carbon samples. Additionally, information regarding Saratech and Carboact is listed for comparison. Our measurement results indicate that the S-4 spherical activated carbon obtained from QITC exhibits promising potential for utilization in low-background experiments.

\begin{table}[h]
	\centering
	\setlength{\tabcolsep}{1.2mm}{
		\footnotesize
		\caption {The $^{226}$Ra concentrations and approximate prices of the six types of activated carbon. S-1, S-2, S-3, S-4, and S-6 are commercially available products, the prices are the approximate prices given by the sales, S-5 is developed by CUP and its price is the small-quantities production cost. }
		
		\begin{tabular}{ccccc}
		\hline
			Sample & $^{226}$Ra concentration (mBq/kg) & Manufactures/Sales & Price (USD/kg) & Ref \\\hline 
			  
			S-1 & 0.23 $\pm$ 0.19  & Carboact & $\sim$25,000 & \citep{r15} \\		
			S-2 & 1.36 $\pm$ 0.94  & Saratech & $\sim$150     & \citep{r13}\\
			S-3 & 49.2 $\pm$ 5.4   & QITC     & $\sim$100    & This work \\
			S-4 & 2.13 $\pm$ 0.41   & QITC     & $\sim$200    & This work\\
			S-5 & 16.0 $\pm$ 1.7   & CUP      & $\sim$150    & This work\\
            S-6 &  400 $\pm$ 41    & SACC     & $\sim$15     & This work \\\hline 
	\label{background}	
	\end{tabular}}
\end{table}

\section*{Radon adsorption capability study}
\subsection*{Measurement device}
\begin{figure}[h]
	\centering
	\includegraphics[width=1\linewidth]{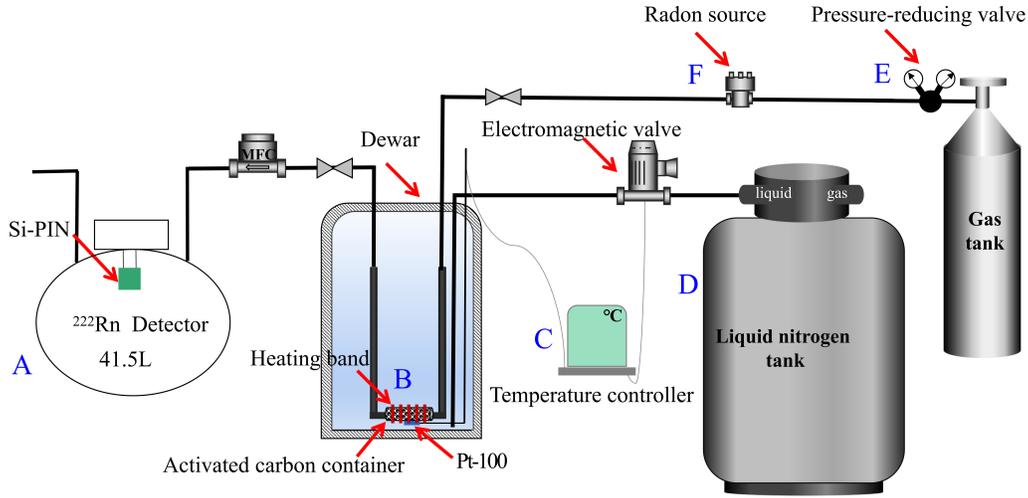}
	\caption{The schematic diagram of the apparatus for the measurement of dynamic adsorption coefficient}
	\label{ACmeasurement}
\end{figure}

In order to investigate the radon adsorption capability of various types of activated carbons, a dedicated system has been developed, as illustrated in Fig.~\ref{ACmeasurement}. In contrast to the devices discussed in Ref~\citep{r14,r35,r37}, our measurement device incorporates a heating system that effectively desorbs the radon gas adsorbed by the activated charcoal at the beginning of the experiment, which could enhance the measurement accuracy. Furthermore, we utilized a highly sensitive radon detector that could exhibit rapid responsiveness to the variations of radon concentration in the gas, thereby minimizing experimental uncertainties. Our system consists of a radon detector, an activated carbon container, a temperature control system, a liquid nitrogen tank, a high-pressure gas cylinder, a mass flow controller (MFC), a radon source, and the associated valves.

(A) The radon detector and the MFC are the same as the one described in Sec.~\ref{emanantionsystem}.

(B) The activated carbon container is a 15~cm long 1/4 VCR pipe equipped with 60 $\mu$m pore size filters at both ends.

(C) The temperature control system consists of a dewar, a Pt100 temperature sensor, a heating belt, an electromagnetic valve, and a temperature controller. The Pt100 temperature sensor is placed between the activated carbon container and the heating belt to serve as the input to the temperature controller. When the set temperature is higher than the present temperature, the temperature controller controls the electromagnetic valve to intermittently inject the liquid nitrogen into the dewar. The liquid nitrogen vaporizes in the dewar tank and the cooled nitrogen cools the activated carbon. When the set temperature is lower than the present temperature, the temperature controller activates the heating belt for heating. With this system, we can achieve temperature regulation from -196~$^{\circ}$C to 200~$^{\circ}$C and the temperature stability is $\sim$5~$^{\circ}$C.

(D) The liquid nitrogen tank here is the same as the one described in Sec.~\ref{emanantionsystem}, but in this case, the piping is connected to the liquid nitrogen outlet for cooling purposes.

(E) The high-pressure cylinder is utilized to supply the required gas for the experiments. The inner pressure of the cylinder is 15~MPa, and a pressure-reducing valve is connected to its outlet to control the gas pressure.

(F) The radon source employed in this system is a gas flow radon source, made from BaRa(CO$_2$)$_3$ powder by University of South China. The radon concentration in the gas is inversely proportional to the gas flow rate. Specifically, when the gas flow rate is set to 3~L/min, the radon concentration in the gas is 175 $\pm$ 18~Bq/m$^3$.

Additionally, knife-edge flanges with metal gaskets and VCR connectors with metal gaskets are used in this system to ensure airtight connections. The air leakage in the system is less than 1 $\times$ 10$^{-8}$ Pa$\cdot$m$^3$/s.

\subsection*{Calculation of radon adsorption coefficient}
\begin{figure}[h]
	\centering
	\includegraphics[width=0.8\linewidth]{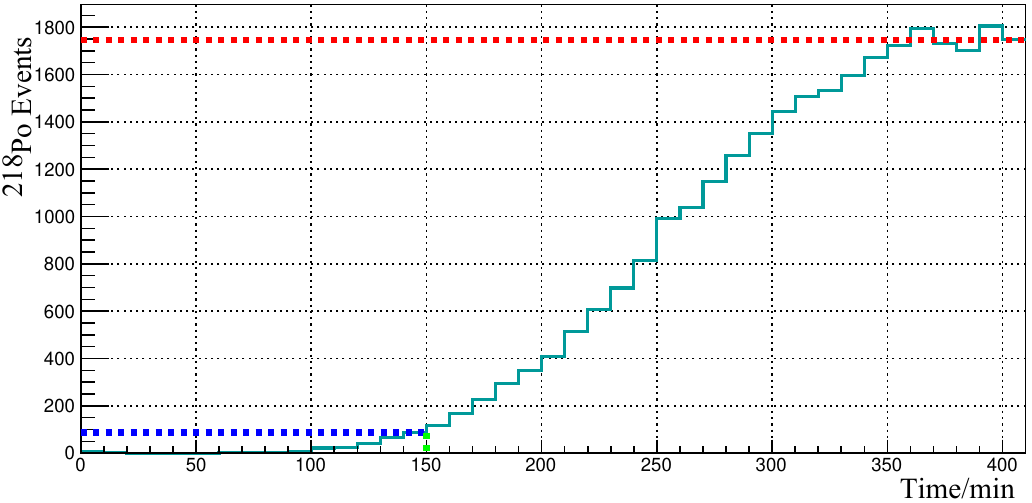}
	\caption{The breakthrough curve of S-4 activated carbon. During the measurement, the temperature is set to -80~$^{\circ}$C, the pressure of the gas is set to 0.4~MPa, and the gas flow rate of nitrogen is set to 3~L/min. The average event rate at equilibrium is indicated with the red dashed line, and the five percent event rate is indicated with the blue dashed line.}
	\label{breakthrough}
\end{figure}

The radon adsorption coefficient is used to quantify the adsorption capability of activated carbon for radon gas. It is influenced by factors such as the temperature of the activated carbon, the type of carrier gas used, and the gas flow rate. Ref.~\citep{r10,r30} provides further details on the relationship between the radon adsorption coefficient and these parameters. To calculate the radon adsorption coefficient, the radon breakthrough curve is typically employed. Fig.~\ref{breakthrough} illustrates an example of a breakthrough curve obtained from an experiment conducted using 0.64 g of S-4 activated carbon. The measurement was performed at a temperature of -80$^{\circ}$C, a pressure of 0.4~MPa, and used evaporated nitrogen as the carrier gas at a flow rate of 3 L/min.

The breakthrough curve is measured as follows:

(A) Adjust the temperature controller to heat the activated carbon to 200~$^{\circ}$C, open the relevant valves after the temperature stabilizes, and purge the system with evaporated nitrogen at a flow rate of 2~L/min for 2~hours. This step helps remove the residual gases or contaminants from the system.

(B) Adjust the temperature controller to cool the activated carbon to -80~$^{\circ}$C. After the temperature is stabilized, adjust the MFC and open the corresponding valves to make the gas flow through the radon source and the activated carbon at a flow rate of 3~L/min.

(C) Monitor the radon concentration in the detector and close the system after the event rate reaches equilibrium.  Once the event rate stabilizes, indicating that the radon breakthrough has occurred, the system can be closed. 

During the measurement of the breakthrough curve, the event rate of $^{218}$Po is used as a proxy for characterizing the radon concentration inside the chamber. This is because $^{218}$Po has a relatively short half-life of 3.1 minutes, which allows it to quickly reflect the radon concentration changes inside the chamber.

The breakthrough curve can reveal the radon adsorption efficiency of the activated carbon. In the first $\sim$100~min, the event rate is nearly zero. This indicates that the activated carbon effectively traps all the radon gas, resulting in a radon adsorption efficiency close to 100\%. During this period, the activated carbon is capable of adsorbing the radon gas efficiently, preventing it from passing through the activated carbon chamber. From $\sim$100~min to $\sim$350~min, the event rate gradually increases. This indicates that the activated carbon is gradually becoming saturated, and its radon adsorption efficiency is progressively decreasing. As the radon gas continues to flow through the activated carbon, some radon molecules manage to bypass the adsorption sites and pass through, leading to an increase in the event rate. After $\sim$350~min, the event rate reaches equilibrium. At this point, the activated carbon becomes fully saturated, and its radon adsorption efficiency drops to zero. The event rate stabilizes, indicating that the activated carbon can no longer effectively adsorb radon gas, and the radon molecules pass through the activated carbon without significant hindrance.

As for the radon adsorption coefficient of activated carbon, different literature has different definitions.  Ref.~\citep{r14,r15,r25} defines it as the volume of gas flowing through the activated carbon when its radon adsorption efficiency drops to zero, Ref.~\citep{r24} uses the volume of gas when the efficiency drops to 50\%, while Ref.~\citep{r13} considers the volume of gas when the efficiency drops to 95\%. Given the specific focus of this paper that the activated carbon is used for radon removal or radon enrichment for low-background experiments, a high radon adsorption efficiency is anticipated. Therefore, we adopt the definition of the radon adsorption coefficient as presented in Ref.~\citep{r13}. The radon adsorption coefficient is calculated according to Eq.~\ref{ADCE}.

\begin{equation} 
	K_{AC} =\frac{F \times t_{0.05}}{m_{AC}} 
\label{ADCE}
\end{equation}

Where K$_{AC}$ is the radon adsorption coefficient in the unit of L/g, F is the gas flow rate in the unit of L/min, and t$_{0.05}$ is the time when the $^{218}$Po event rate reaches 5\% of the equilibrium event rate in the unit of min. m$_{AC}$ is the mass of the activated carbon in the unit of g. As can be seen from Fig.~\ref{breakthrough}, the radon coefficient of S-4 activated carbon at -80~$^\circ$C is 703 $\pm$ 23~L/g, the error is derived from the uncertainty of t$_{0.05}$, which is half of the bin width in Fig.~\ref{breakthrough}.

\subsection*{Effect of pressure on radon adsorption capability}

\begin{figure}[H]
	\centering
	\includegraphics[width=0.8\linewidth]{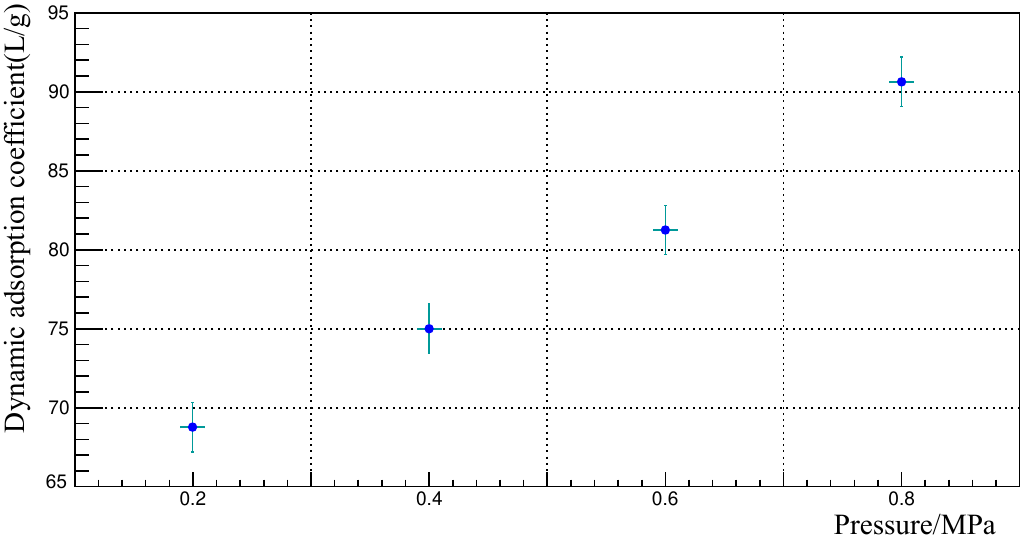}
	\caption{Relationship between dynamic adsorption coefficient and gas pressure.}
	\label{pressure}
\end{figure}

In addition to temperature and gas flow rate, pressure is another factor that can impact the radon adsorption capability of activated carbon. To ensure a fair comparison of the radon adsorption capacity among different types of activated carbon, it is essential to maintain consistent external conditions during the experiments. To investigate the influence of pressure on radon adsorption capability, a test was conducted using 0.64~g of S-4 activated carbon with the experimental setup depicted in Fig.~\ref{ACmeasurement}.

Throughout the test, the temperature was maintained at -40~$^\circ$C, and evaporated nitrogen was employed as the carrier gas. The gas flow rate was set to 1~L/min. This specific flow rate was chosen due to the relatively high flow resistance of the activated carbon chamber, 1~L/min is the maximum flow rate achievable at a pressure of 0.2~MPa. The temperature of -40 $^\circ$C was selected to expedite the experimental duration.

The measurement results, as depicted in Fig.~\ref{pressure}, reveal that the radon adsorption capability of activated carbon increases with increasing pressure. However, it is important to note that the experiment did not test for higher pressures beyond the maximum pressure limit of the liquid nitrogen tank, which is 0.8 MPa. Therefore, the observed trend in radon adsorption capability is limited to the tested pressure range.

\subsection*{Comparasion of radon adsorption capability}
In low-background experiments, both the intrinsic background and the radon adsorption capability are crucial factors affecting the application of activated carbon. In our investigation, we have identified that the S-4 activated carbon from QITC exhibits the lowest background among the tested activated carbons other than Carboact and Saratech activated carbon. To assess its potential application in low-background experiments, we conducted comparative tests of their radon adsorption capacities. During the tests, the temperature is set to -80~$^\circ$C, the nitrogen flow rate is set to 3~L/min, and the pressure is set to 0.4~MPa. 

\begin{figure}[H]
	\centering
	\includegraphics[width=0.8\linewidth]{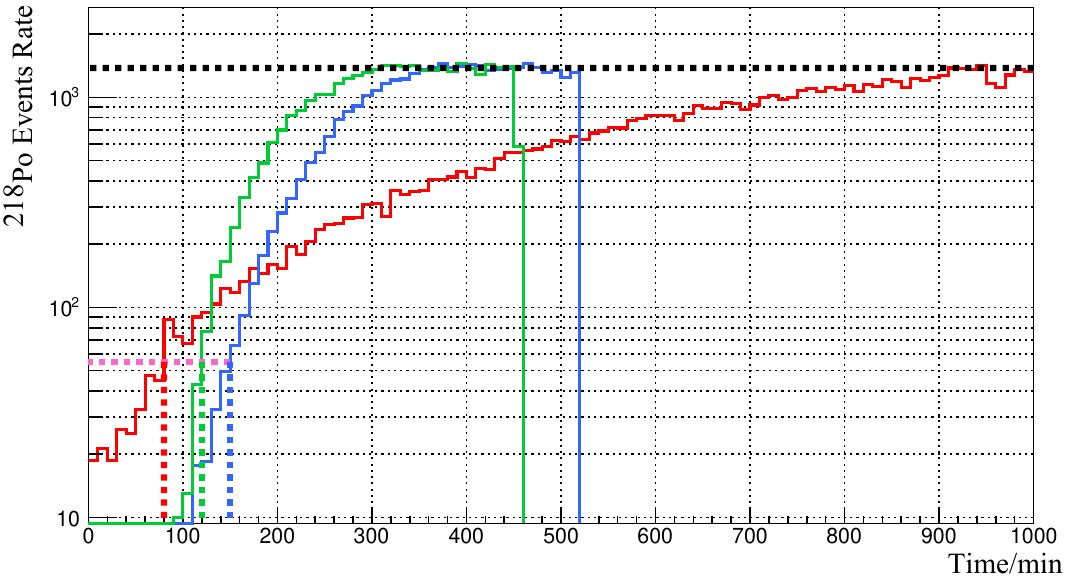}
	\caption{The breakthrough curves of S-1, S-2 and S-4 activated carbons.The red solid curve represents the breakthrough curve of S-1 activated carbon, the blue solid curve represents S-4 activated carbon, and the green solid curve represents S-2 activated carbon. The black dashed line indicates the $^{218}$Po event rate at equilibrium, the pink dashed line indicates the $^{218}$Po event rate when the adsorption efficiency decrease to 95\%, the red, green, and blue dashed lines indicate the times the adsorption efficiency drops to 95\% for the three kinds of activated carbon. }
	\label{breakthroughcurves}
\end{figure}

\begin{table}[h]
	\centering
	\setlength{\tabcolsep}{2mm}{
		\footnotesize
		\caption{Comparasion of radon adsorption coefficient of different kinds of activated carbon. The errors of t$_{0.05}$ are the bin width of the breakthrough curve, and the errors of the adsorption coefficient are derived from the errors of t$_{0.05}$.}
		
		\begin{tabular}{cccc}
			\hline
			Sample & Mass(g) & t$_{0.05}$(min) & Adsorption coefficient (L/g) \\\hline 
			S-1 & 0.88 & 80 $\pm$ 5  & 272 $\pm$ 17    \\
			S-2 & 0.73 & 120 $\pm$ 5 & 493 $\pm$ 21  \\		
			S-4 & 0.64 & 150 $\pm$ 5 & 703 $\pm$ 23\\\hline 
	\label{CofAC}		
	\end{tabular}}
\end{table}

The breakthrough curves are shown in Fig.~\ref{breakthroughcurves} and the relevant measurement information is summarized in Tab.~\ref{CofAC}. From the analysis of the breakthrough curves, it is evident that the S-4 activated carbon from QITC exhibits significantly stronger radon adsorption capability compared to Saratech and Carboact activated carbons. This finding indicates that S-4 activated carbon is an excellent choice as a radon adsorbent for low-background experiments.

\subsection*{Optimization of activated carbon pore size}

The adsorption performance of activated carbon depends on its physical and chemical structures~\citep{r26}. The surface area and the pore size of activated carbon are two critical physical factors that affect the adsorption capability of activated carbon. In order to find higher radon adsorption capability low-background activated carbon, we purchased three other batches of activated carbons from QITC with the same manufacturing technique and raw materials but different surface areas and different pore sizes. The radon adsorption coefficients have been measured with the system shown in Fig.~\ref{ACmeasurement}. The information about the four batches of activated carbon and their radon adsorption capacity are summarized in Tab.~\ref{QITC-4}. According to Tab.~\ref{QITC-4}, S-8 activated carbon has the strongest radon adsorption capability. It is believed that the larger the surface area of activated carbon, the stronger its adsorption capacity. Therefore, this result indicates that the radon adsorption capacity of activated carbon should be more related to its pore size.

\begin{table}[h]
	\centering
	\setlength{\tabcolsep}{1.2mm}{
		\footnotesize
		\caption{The physical factors and radon adsorption coefficients of four batches of activated carbon from QITC. They are made from the same raw material and with the same manufacturing technique.}
\begin{tabular}{ccccc}
			\hline
			Sample & Surface area (m$^2$/g) & Average pore size (nm) & Mass (g) & Adsorption coefficient (L/g) \\\hline 
			S-4 &  1001 & 1.6  & 0.64 &   703 $\pm$ 23  \\
			S-7 &  854  & 1.8 &  0.67 & 940 $\pm$ 22 \\		
			S-8 &  810 & 2.3 & 0.65 & 1315 $\pm$ 21\\
            S-9 &  1096 & 3.0 & 0.63 & 905 $\pm$ 24 \\\hline 
	\label{QITC-4}		
	\end{tabular}}
\end{table}

\subsection*{Radon adsorption capability with different carrier gases}
The measurements mentioned above were conducted using evaporated nitrogen as the carrier gas. However, noble gases are widely employed in low-background experiments~\citep{r28}, and radon is one of the primary background sources~\citep{r27,r29}. To investigate the adsorption capacity of the activated carbon for radon gas in different noble gases, we measured the radon adsorption coefficients using argon, neon, and krypton as carrier gases. These noble gases were sourced from high-pressure gas cylinders. The measurement setup, as depicted in Fig.~\ref{ACmeasurement}, utilized 0.65 g of S-8 activated carbon. The measurements were performed under the following conditions: a temperature of -80 $^\circ$C, a gas flow rate of 3 L/min, and a pressure of 0.4 MPa. The aerodynamic equivalent diameters of the atoms and the measured radon adsorption coefficients are presented in Table~\ref{dcg}.

As observed in Table~\ref{dcg}, the radon adsorption capability decreases as the diameter of the carrier gas increases. This can be attributed to the fact that activated carbon can adsorb gases other than radon. The activated carbon used in this study was optimized for radon adsorption, which has the largest atomic diameter among the gases used. As the atomic diameter increases, the adsorption capacity of activated carbon also increases. Consequently, the adsorption sites within the activated carbon become occupied by the carrier gas atoms, leading to a gradual decrease in its adsorption capacity for radon. This phenomenon also explains the challenges associated with removing radon from xenon gas using activated carbon.

\begin{table}[H]
	\centering
	\setlength{\tabcolsep}{2mm}{
		\footnotesize
		\caption{ Dynamic adsorption coefficients of S-C under different carrier gases }
		
		\begin{tabular}{ccc}
			
			\hline
		Carrier gas & Dynamic adsorption coefficient (L/g) & Aerodynamic equivalent diameter(Å) \\\hline 
			Ne &  2014 $\pm$ 21  &   3.08  \\
			Ar &   891 $\pm$ 23   &   3.40  \\
			Kr &  42.0 $\pm$ 2.3   &   3.67 \\
            Rn &     --       &   4.17 \\\hline 
    \label{dcg}
	\end{tabular}}
\end{table}

\section*{Conclusion}

Activated carbon is widely used as an efficient adsorbent for radon in low-background experiments, including neutrino research and dark matter detection, where radon and its decay products are significant background sources. The radon adsorption capacity of activated carbon and its intrinsic $^{226}$Ra concentration are crucial factors in these experiments. In this paper, we developed a 73~$\mu$Bq/sample radon emanation measurement system, measured the $^{226}$Ra concentrations of four different types of activated carbon with this system, and obtained a spherical activated carbon with a $^{226}$Ra concentration of 2.13 $\pm$ 0.41~mBq/kg.

Using a self-developed apparatus, we measured the radon adsorption coefficient of the activated carbon and found that its radon adsorption capacity was significantly higher than that of Saratech or Carboact activated carbons under the same conditions. Furthermore, when the average pore size of this activated carbon was adjusted to 2.3~nm, its radon adsorption capacity was further enhanced. Under conditions of -80~$^\circ$C, 0.4~MPa pressure, and a gas flow rate of 3~L/min, this activated carbon exhibited a radon adsorption capacity of 1315 $\pm$ 21~L/g with nitrogen as the carrier gas, which is approximately 2.7 times or 4.8 times of Saratech or Carboact activated carbon. We also measured the radon adsorption coefficients of S-8 activated carbon with different noble carrier gases, the results showed that as the increase of carrier gases' atomic diameter, its radon adsorb capability decreases.

According to the measurement results, S-8 spherical activated carbon from QITC company is an excellent candidate for radon removal and radon enrichment in low-background experiments.


\section*{Declaration of competing interest}
There are no conflicts to declare.

\section*{Data availability}
Data will be made available on request.

\section*{Acknowledgement}
This work is supported by the State Key Laboratory of Particle Detection and Electronics (Grant No. SKLPDE-ZZ-202304), the Youth Innovation Promotion Association of Chinese Academy of Sciences, and the Yalong River Joint Fund of the National Natural Science Foundation of China and Yalong River Hydropower Development Co., LTD (Grant No. U1865208).



\begin{thebibliography}{00}
\bibitem[JUNO Collaboration(2022)]{r1}JUNO Collaboration, JUNO physics and detector, PPNP 123, 2022, 103927. https://doi.org/10.1016/j.ppnp.2021.103927.
\bibitem[Alimonti et al.(2009)]{r2}Alimonti, G., Arpesella, C., et al., The Borexino detector at the Laboratori Nazionali del Gran Sasso, NIM A 600  2009  568-593. https://doi.org/10.1016/j.nima.2008.11.076.
\bibitem[Fukuda et al.(2003)]{r3}Fukuda, S., Fukuda, Y., et al., The Super-Kamiokande detector, NIM A 501 2003 418-462. https://doi.org/10.1016/S0168-9002(03)00425-X.
\bibitem[Aprile et al.(2017)]{r4}Aprile, E., Aalbers, J., et al., The XENON1T dark matter experiment, EPJC 12 2017 881. https://doi.org/10.1140/epjc/s10052-017-5326-3.
\bibitem[Bonet et al.(2021)]{r5}Bonet, H., Bonhomme, A., et al., Constraints on Elastic Neutrino Nucleus Scattering in the Fully Coherent Regime from the CONUS Experiment, PRL 126 2021 041804. https://doi.org/10.1103/PhysRevLett.126.041804
\bibitem[Guo et al.(2020)]{r6}Guo, C., Guan, M.Y., et al., The liquid argon detector and measurement of SiPM array at liquid argon temperature, NIM A 980 2020 164488. https://doi.org/10.1016/j.nima.2020.164488.
\bibitem[Baudis et al.(2018)]{r7}Baudis, L., Biondiet, Y., al., A dual-phase xenon tpc for scintillation and ionization yield measurements in liquid xenon, EPJC 78 2018 351. https://doi.org/10.1140/epjc/s10052-018-5801-5.
\bibitem[Akimov et al.(2018)]{r8}Akimov, D., Albert, J.B., et al., COHERENT 2018 at the Spallation Neutron Source. https://doi.org/10.48550/arXiv.1803.09183.
\bibitem[Angloher et al.(2021)]{r9}Angloher, G., Bharadwaj, M.R., et al., First measurements of remoTES cryogenic calorimeters: Easy-to-fabricate particle detectors for a wide choice of target materials, NIM A 1045 2021 167532. https://doi.org/10.1016/j.nima.2022.167532.
\bibitem[Aprile et al.(2021)]{r18}Aprile, E., Aalbers, J., et al., $^{222}$Rn emanation measurements for the XENON1T experiment, EPJC 81 2021 337. https://doi.org/10.1140/epjc/s10052-020-08777-z

\bibitem[Liu et al.(2022)]{r10}Liu, Y., Zhang, Y.P., Liu, J.C., et al., System upgrade for $\mu$Bq/m$^{3}$ level $^{222}$Rn concentration measurement. JINST 18 2022 T03002. https://doi.org/10.1088/1748-0221/18/03/T03002.
\bibitem[Simgen, Zuzel.(2009)]{r11}Simgen, H., Zuzel, G., Analysis of the $^{222}$Rn concentration in argon and a purification technique for gaseous and liquid argon, Appl. Radiat. Isot. 67 2009 922-925. https://doi.org/10.1016/j.apradiso.2009.01.058.
\bibitem[Abe et al.(2012)]{r12}Abe, K., Hieda, K., et al., Radon removal from gaseous xenon with activated charcoal, NIM A 661 2012 50-57. https://doi.org/10.1016/j.nima.2011.09.051.
\bibitem[Chen et al.(2022)]{r13}Chen, Y.Y., Zhang, Y.P., Liu, Y., et al., A study on the radon removal performance of low background activated carbon, JINST 17 2022 P02003. https://doi.org/10.1088/1748-0221/17/02/P02003.
\bibitem[Guo et al.(2017)]{r14}Guo, L., Wang, Y., Zhang, L., Zeng, Z., Dong, W., Guo, Q., The temperature dependence of adsorption coefficients of $^{222}$Rn on activated charcoal: an experimental study, Appl. Radiat. Isot. 125 2017 185–187. http://dx.doi.org/10.1016/j.apradiso.2017.04.023.
\bibitem[Pushkin et al.(2018)]{r15}Pushkin, K., Akerlof, C., et al., Study of radon reduction in gases for rare event search experiments, NIM A 903 2018 267-276. https://doi.org/10.1016/j.nima.2018.06.076.
\bibitem[Wojcik et al.(2017)]{r16}Wojcik, M., Zuzel, G., and Simgen, H., Review of high-sensitivity Radon studies, Int. J. Mod. Phys. 32 2017 1743004. https://doi.org/10.1142/S0217751X17430047.

\bibitem[Saratech(2023)]{r19}Saratech charcoal specification, https://www.bluecher.com/en/. (Accessed 13 July 2023). Last time accessed.
\bibitem[Carboact(2023)]{r21}Carboact charcoal specification, https://www.carboactinternational.com/. (Accessed 13 July 2023). Last time accessed.
\bibitem[Nakano et al.(2017)]{r22}Nakano, Y., Sekiya, H., Tasaka, S., Takeuchi, Y., Wendell, R.A., Matsubara, M., et al., Measurement of radon concentration in super-Kamiokande’s buffer gas, NIM A 867 2017 108-114. http://dx.doi.org/10.1016/j.nima.2017.04.037.
\bibitem[Karunakara et al.(2015)]{r23}Karunakara, N., Sudeep Kumara, K., Yashodhara, I.,  Sahoo, B.K., Gaware, J.J., Sapra, B.K., Mayya, Y.S., Evaluation of radon adsorption characteristics of a coconut shell-based activated charcoal system for radon and thoron removal applications, Journal of Environmental Radioactivity. 142 2015 87-95. http://dx.doi.org/10.1016/j.jenvrad.2014.12.017.
\bibitem[Li et al.(2022)]{r24}Li, C., Tang, Q., Feng, X. et al., Construction of a low-temperature activated carbon radon adsorption system using air cooler, J. Radioanal. Nucl. Chem. 331 2022 1839-1845. 
https://doi.org/10.1007/s10967-022-08270-9.
\bibitem[Fan Wang et al.(2023)]{r25}Fan Wang, Hao Wang, Hao Ma, Lei Zhang, Changhao Sun, Qiuju Guo, Radon dynamic adsorption coefficients of two activated charcoals at different temperatures in nitrogen environment, Appl. Radiat. Isot. 191 2023 110564. https://doi.org/10.1016/j.apradiso.2022.110564.
\bibitem[Wang et al.(2011)]{r26}Wang, Q., Qu, J., Zhu, W., Zhou, B., Cheng, J., An experimental study on radon adsorption ability and microstructure of activated carbon, Nuclear science and engineering. 168 2021 287-292. http://dx.doi.org/10.13182/NSE10-65.
\bibitem[Rupp(2017)]{r27}Rupp, N., Radon background in liquid xenon detectors, JINST 13 2017 C02001. https://doi.org/10.1088/1748-0221/13/02/C02001.
\bibitem[Aalseth et al.(2020)]{r28}Aalseth, C.E., Abdelhakim, S., et al., Design and construction of a new detector to measure ultra-low radioactive-isotope contamination of argon, JINST 15 2020 P02024. https://doi.org/10.1088/1748-0221/15/02/P02024.
\bibitem[Agnes et al.(2019)]{r29}Agnes, P., Albuquerque, I.F.M., et al., Measurement of the ion fraction and mobility of $^{218}$Po produced in $^{222}$Rn decays in liquid argon, JINST 14 2019 P11018. https://doi.org/10.1088/1748-0221/14/11/P11018.
\bibitem[Qiao et al.(2011)]{r30}Qiao, B., Liu, C., Zheng, W., et al., Dynamic adsorption properties of activated carbon for radioactive noble gas treatment in offshore floating nuclear power plant, J. Radioanal. Nucl. Chem. 327 2021 207–215. https://doi.org/10.1007/s10967-020-07382-4.
\bibitem[G. Heusser et al. (2000)]{r31} G. Heusser a, et al., $^{222}$Rn detection at the $\mu$Bq/m$^3$ range in nitrogen gas and a new Rn purification technique for liquid nitrogen, Applied Radiation and Isotopes 52 (2000) 691–695
\bibitem[X. Yu et al. (2020)]{r32} X. Yu, et al., Radon activity measurement of JUNO nitrogen, JINST 15  (2020) P09001
\bibitem[Stephen Robert Hanchurak (2014)] {r33} Stephen Robert Hanchurak, Development of a high sensitivity radon emanation detector, Master thesis, Department of Physics, University of Alberta
\bibitem[B. Soule (2013)]{r34} B. Soule, Radon emanation chamber: High sensitivity measurements for the SuperNEMO experiment, AIP Conference Proceedings 1549 (2013) 98
\bibitem[Y. Nakano (2020)]{r35} Y. Nakano, et al., Evaluation of radon adsorption efficiency values in xenon with activated carbon fibers, Prog. Theor. Exp. Phys. 2020, 113H01
\bibitem[K. Pushkin (2018)]{r37} K. Pushkin, et al., Study of radon reduction in gases for rare event search experiments, NIM A 903 (2018) 267-276
\end{thebibliography}

\end{document}